**Pressure-induced structural transition, metallization, and topological superconductivity in PdSSe**

Feng Xiao[1], Wen Lei[1,2], Wei Wang[1,3], Carmine Autieri[4], Xiaojun Zheng[1], Xing Ming[1*], and Jianlin Luo[5,6]

1. College of Science, MOE Key Laboratory of New Processing Technology for Nonferrous Metal and Materials, Guilin University of Technology, Guilin 541004, PR China

2. Key Laboratory of Artificial Micro- and Nano-Structures of Ministry of Education and School of Physics and Technology, Wuhan University, Wuhan 430072, PR China

3. School of Materials Science and Engineering, Beihang University, Beijing 100191, PR China

4. International Research Centre MagTop, Institute of Physics, Polish Academy of Sciences, Aleja Lotników 32/46, PL-02668 Warsaw, Poland

5. Beijing National Laboratory for Condensed Matter Physics and Institute of Physics, Chinese Academy of Sciences, Beijing 100190, PR China

6. School of Physical Sciences, University of Chinese Academy of Sciences, Beijing 100190, PR China

---

[*] Email: mingxing@glut.edu.cn (Xing Ming)

**ABSTRACT**

Pressure not only provides a powerful way to tune the crystal structure of transition metal dichalcogenides (TMDCs) but also promotes the discovery of exotic electronic states and intriguing phenomena. Structural transitions from the quasi-two-dimensional layered orthorhombic phase to three-dimensional cubic pyrite phase, metallization, and superconductivity under high pressure have been observed experimentally in TMDCs materials $PdS_2$ and $PdSe_2$. Here, we report a theoretical prediction of the pressure-induced evolutions of crystal structure and electronic structure of PdSSe, an isomorphous intermediate material of the orthorhombic $PdS_2$ and $PdSe_2$. A series of pressure-induced structural phase transitions from the layered orthorhombic structure into an intermediate phase, then to a cubic phase are revealed. The intermediate phase features the same structure symmetry as the ambient orthorhombic phase, except for drastic collapsed interlayer distances and striking changes of the coordination polyhedron. Furthermore, the structural phase transitions are accompanied by electronic structure variations from semiconductor to semimetal, which are attributed to bandwidth broaden and orbital-selective mechanisms. Especially, the cubic phase PdSSe is distinct from the cubic $PdS_2$ and $PdSe_2$ materials by breaking inversion and mirror-plane symmetries, but showing similar superconductivity under high pressure, which is originated from strong electron-phonon coupling interactions concomitant with topologically nontrivial Weyl and high-fold Fermions. The intricate interplay between lattice, charge, and orbital degrees of freedom as well as the topologically nontrivial states in these compounds will further stimulate wide interest to explore the exotic physics of the TMDCs materials.

# I. INTRODUCTION

Since the discovery of atomic thin graphene, two-dimensional (2D) materials have received extensive attention [1,2]. As one important branch of 2D materials, transition metal dichalcogenides (TMDCs) materials have attracted more and more interest due to their superior electronic, optical, mechanical, and catalytic properties [3]. Layered TMDCs materials $MX_2$ are commonly composed of quasi-2D $X$-$M$-$X$ type sandwich layers held together by weak van der Waals (vdW) forces, in which $M$ is the transition metal element and $X$ is chalcogen element such as S, Se, or Te. As a new member of the TMDCs family, the noble-metal dichalcogenides (NMDCs) Pd$X_2$ materials are potential candidates for exotic physical properties and novel technical applications at the nanoscale [4-6]. Among them, PdTe$_2$ tends to crystallize in a $1T$ structure with octahedral coordination [7,8], whereas PdS$_2$ and PdSe$_2$ adopt an orthorhombic structure named as $2O$ phase at ambient conditions [9,10], in which each Pd atom coordinates with four chalcogen atoms in a $(PdX_4)^{2-}$ square plane and results in a puckered pentagonal morphology. Due to the strange pentagonal structure and unusual interlayer coupling interactions [11,12], the $2O$ phase PdS$_2$ and PdSe$_2$ materials host excellent physical-chemical properties and promising application prospects [4-6], including layer-dependent tunable band gaps [13-16], highly anisotropic properties [17-26], and ultra-high air stability [16], which are promising materials for field-effect transistor [17,23], broadband photodetector [18,27,28], thermoelectric applications [13,20-22], and nonlinear photoelectronic devices [29].

Similar to conventional 2D semiconductors, the physical properties of Pd$X_2$ are susceptible to the external perturbations, such as light [29], electric field [30], molecular adsorption [31], applied pressure [32-36], and strain [24,26,34,37-39]. Particularly, pressure and strain are two effective means to tailor the crystal structures and electronic structures of Pd$X_2$. PdS$_2$ and PdSe$_2$ undergo structure transitions from the layered $2O$ phase to cubic pyrite-type structure under pressure, accompanied by metallization and superconductivity [32-35]. Theoretical investigations propose that the metallic electron structure originates from an orbital-selective mechanism [32,34]. High pressure experiments reveal a dome-shaped relationship of the superconducting temperature with the pressure for the high-pressure pyrite phase PdS$_2$ and PdSe$_2$ [32,35]. Electronic structure calculations indicate that the superconductivity originates from strong electron-phonon coupling interactions combined with topological nodal-line states [34,35]. Meanwhile, excellent ferroelasticity with a low energy barrier and strong switching signal under strain has been theoretically predicted for PdS$_2$ and PdSe$_2$ with great potential for the application of shape memory devices [24,26,34,37]. Furthermore, these ferroelastic transitions accompanied with crystal orientation rotations give rise to tunable transport properties and switchable thermal conductivity [24,26,34,38].

Not only external stimuli, but also internal factors such as vacancy defects [15,40-43], chemical doping [44-46], and surface oxidation [47] can be utilized to trigger structure transitions and manipulate the physical-chemical properties of $PdS_2$ and $PdSe_2$. Experiments demonstrate that plasma irradiation [40] or electron irradiation [42] can induce Se atom vacancy in $PdSe_2$, which breaks the Se-Se dumbbell-shaped dimers and results in huge structural distortions and formation of metallic $Pd_{17}Se_{15}$ and novel 2D $Pd_2Se_3$ materials [40,42]. The defect states in $PdSe_2$ can be reversibly switched between charge neutral and charge negative, and show anisotropic carrier mobility [41,43]. First-principles theoretical investigations show that hole doping can induce half-metallic ferromagnetism in the $PdSe_2$ monolayer, and the magnetization direction can be controlled by a transverse uniaxial stress [46]. Furthermore, controllable partial oxidation of the surface of $PdSe_2$ by ozone treatment successfully modifies its transport property, optoelectronic performance, and catalytic activity [47]. On the other hand, due to the different ionic radii, doping or substitution can induce chemical pressure in crystals, which are considered as specific methods conducive to stabilizing pressure-induced superconductivity or other characteristic phenomena at ambient conditions [48]. Recently, Te substitution induced structural evolutions from the orthorhombic 2$O$ phase to a monoclinic Verbeekite phase and then to a trigonal 1$T$ phase are observed in the $PdSe_{2-x}Te_x$ system, which are accompanied by a drastic change of the electrical transport behavior from semiconductor to metal and enhanced superconductivity with increasing Te content [44,49]. In addition, a solid solution of $PdSe_{2-x}S_x$ with a wide range of $x$ ($0 \leq x \leq 0.86$) has been synthesized by atomic substitution, which shows tunable density of states, effective mass, and band gap by changing $x$ [45].

Although 2$O$ phase PdSSe, an isomorphous intermediate material of orthorhombic $PdS_2$ and $PdSe_2$, has been synthesized for over half a century [10], its detailed crystal structure is only reported by our theoretical works recently [50]. As shown in **Figure 1**, the bulk phase of PdSSe is constituted by two types of monolayer PdSSe (denoted as $A$ and $B$) and forms three polymorphs with different stacking manners (denoted as $AA$, $AB$, and $BB$). Previous theoretical predictions indicate that PdSSe will exhibit intriguing properties including excellent photocatalytic properties and tunable band gap under biaxial strain, which are promising candidates for high-performance optoelectronic, nanoelectronic, photovoltaic and photocatalytic applications [50,51]. Although tremendous focus has been paid to $PdS_2$ and $PdSe_2$, the study on their counterpart PdSSe is scarce. Considering that $PdS_2$ and $PdSe_2$ are featured with rich phase transitions, tunable properties and interesting superconducting behavior under stress or pressure [24,26,32-39], one may naturally expect similar behavior to be observed in their sister material PdSSe. In the present work, we explored the structural evolutions of the bulk phase PdSSe under hydrostatic pressure, uncovered an electronic structure transition from semiconductor to semimetal and superconductor, revealed the origin of the metallization, and predicted topological superconductivity for the high-pressure phase. The predicted structural transitions and unconventional transport properties concomitant with topologically nontrivial states in PdSSe supply a meaningful complement to the hectic NMDCs material of PdSe$_2$ and PdS$_2$, which will attract extensive interest from the wide audience to explore its unanticipated properties.

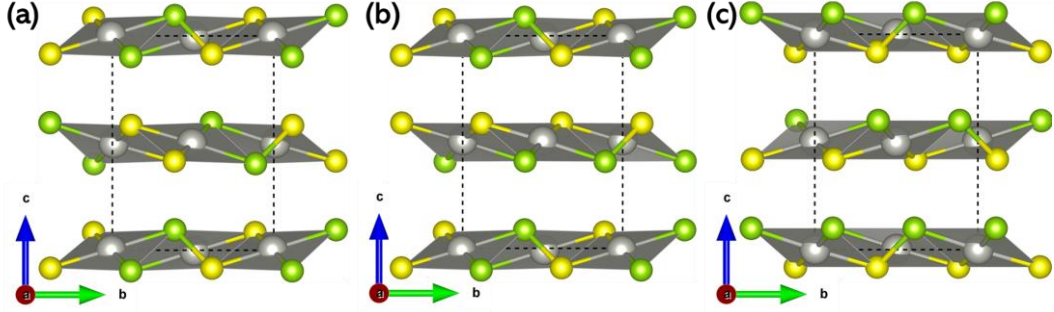

**Figure 1** Three polymorphs of PdSSe in (a) *AA*, (b) *AB*, and (c) *BB* stacking manner with gradually increasing energy. The big gray balls represent Pd atoms, while the small yellow (green) balls represent S (Se) atoms.

## II. CALCULATION METHODS

Geometry optimization and electronic structure calculations were performed by using density functional theory as implemented in the Vienna *Ab initio* Simulation Package (VASP) [52,53]. The projector-augmented-wave potential, the Perdew-Burke-Ernzerhof (PBE) exchange-correlation functional [54], and the plane-wave basis set with a cutoff energy of 520 eV were employed. The smallest allowed spacing of the reciprocal space between *k* points was fixed to 0.2 Å$^{-1}$. In order to obtain a more accurate band gap close to the experimental value, Heyd-Scuseria-Ernzerhof (HSE06) [55,56] hybrid density functional was adopted to calculate the band structures. Phonon calculations were performed by the density functional perturbation theory with the QUANTUM-ESPRESSO codes [57]. Electron-phonon coupling constants were calculated on phonon and electron grids of 4 × 4 × 4 and 16 × 16 × 16, respectively, with Gaussians smearing width of 0.05 Ry. The plane-wave cutoff energy was set to 50 Ry for all three polymorphs of PdSSe.

The selection of the vdW correction scheme is significant for the study of layered vdW materials by first-principles calculations. Due to the lack of a detailed crystal structure of PdSSe, the choice of the vdW scheme was based on its sister compounds PdS$_2$ and PdSe$_2$ in our previous work [50]. In the present work, further screening was carried out, and dDsC dispersion correction [58,59] was adopted to consider the interlayer vdW interactions [**Table S1** in the Supplemental Material [60]. As shown in **Table 1**, the theoretically optimized lattice constants are closer to the experimental value than our previous results (Table 1 in Ref. [50]), demonstrating the validity of the calculation parameters. Additionally, the energies are comparable with each other for these three polymorphs of PdSSe, and they show similar structural phase transition and high-pressure superconductivity. Therefore, we mainly presented the results and discussions of the *AA* polymorph of PdSSe in the main text, and the results for the other two polymorphs (*AB* and *BB*) were provided in the Supplemental Material [60].

**Table 1** The optimized lattice constants, unit cell volume ($V_0$), and relative energy $E_r$ of the three most stable polymorphs of bulk phase PdSSe compared with the experimental results.

|       | *a* (Å) | *b* (Å) | *c* (Å) | $V_0$ (Å$^3$) | $E_r$ (meV/atom) |
|-------|---------|---------|---------|---------------|------------------|
| Expt.[10] | 5.595 | 5.713 | 7.672 | 245.23 | - |
| *AA*  | 5.628 | 5.742 | 7.663 | 247.665 | 0.00 |
| *AB*  | 5.631 | 5.740 | 7.664 | 247.718 | 1.60 |
| *BB*  | 5.639 | 5.743 | 7.625 | 246.951 | 1.79 |

## III. RESULTS AND DISCUSSION

### A. Structural evolution under hydrostatic pressure

The structural evolution under compression has been simulated by gradually increasing the pressure from 0 GPa to 10 GPa. As shown in **Figure 2(a)**, the lattice constants show an anisotropic response to the hydrostatic pressure. Attributed to the weak interlayer connection of vdW bonds, the out-of-plane lattice constant *c* significantly shrinks when pressure increases to 3 GPa. In contrast, the lattice constants *a* and *b* increase slowly. The vdW radii of Pd, S, and Se ions are 2.15, 1.8, and 1.9 Å, whereas the covalent radii of Pd, S, and Se ions are 1.39, 1.05 and 1.20 Å [61,62]. Apparently, the interlayer and intralayer Pd-*X* bonds belong to vdW and covalent bonds, respectively. Compared with the easily compressed interlayer vdW bonds along the stacking direction, the in-plane intralayer covalent bonds in the (Pd-S$_2$-Se$_2$)$^{2-}$ square-plane are less affected by applied pressure [**Figure 2 (b)**]. At a pressure of 3.5 GPa, the orthorhombic PdSSe transforms to an intermediate phase. The intermediate phase still has the identical crystallographic symmetry with the ambient-pressure phase, but the out-of-plane lattice constants *c* dramatically shrinks accompanied by obvious expansions of the in-plane lattice constants *a* and *b*. Meanwhile, the interlayer Pd-*X* distances between Pd atoms and chalcogen atoms *X* in the adjacent layer decrease sharply, which are close to the intralayer counterparts. This drastic structural change causes a sudden volume collapse [**Figure 2 (c)**] and energy increasing [**Figure 2 (d)**]. All the lattice constants become equal and the orthorhombic PdSSe transforms to a cubic phase above 4 GPa (the structure symmetry will discussed later in the subsection of **II. C**). Furthermore, three distinct stages of structural transition in the isostructural and isoelectronic $PdS_2$ and $PdSe_2$ have been observed by high pressure experiments, the 2$O$ orthorhombic phase has been proposed to coexist with the cubic pyrite phase in the intermediate stage [32,33,35]. Our theoretical calculations indicate that an intermediate phase may exist in the structural transition process, which implies that a coexisting region of the orthorhombic phase with the cubic phase may be observed in future high-pressure experiments.

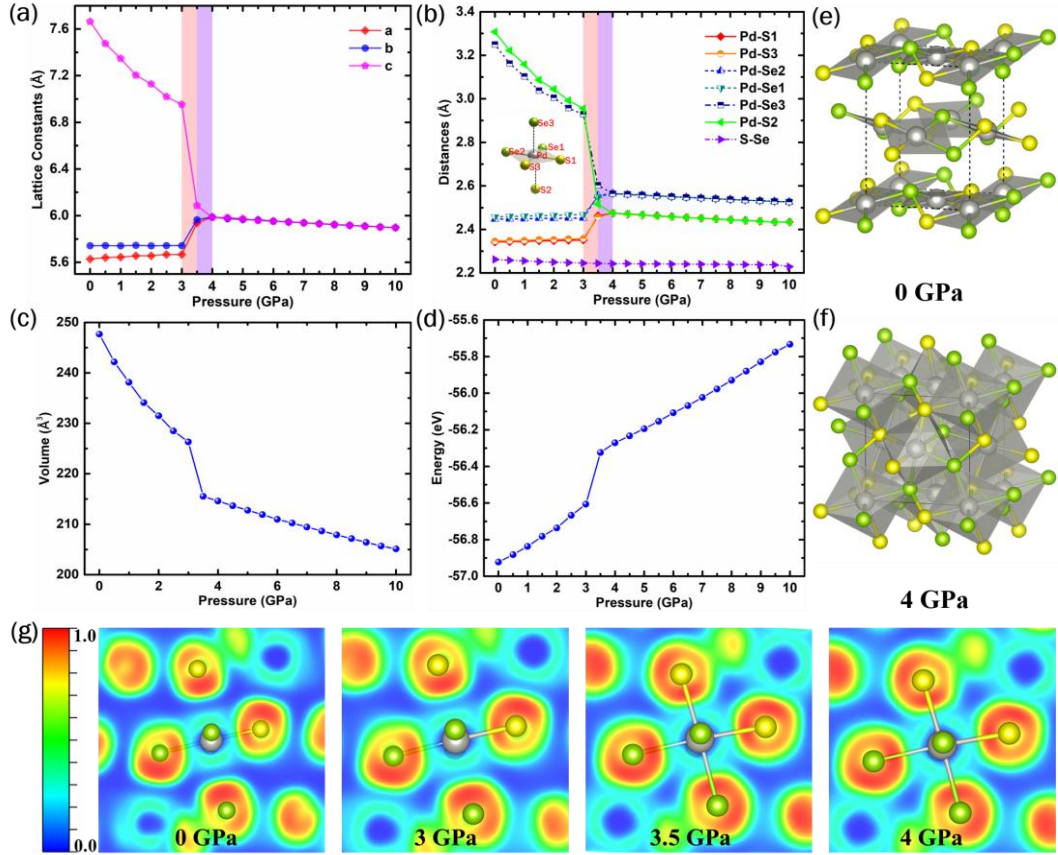

**Figure 2** Evolutions of crystal structure and electronic structure of PdSSe under hydrostatic pressure: (a) lattice constants, (b) interatomic distances (the inset illustrates the coordination environment of the Pd atom), (c) volume, and (d) energy changes as a function of pressure, polyhedral crystal structure at (e) 0 GPa and (f) 4 GPa, and (g) slices of the electron localization function at different pressures.

According to the pressure homology principle, the origin of the pressure-induced structural transition of PdSSe from the 2$O$ orthorhombic phase to the cubic phase can be explained by the displacive mechanism [33]. In fact, the layered orthorhombic phase is derived from a structural distortion of the cubic phase, and the unusual $(Pd-S_2-Se_2)^{2-}$ square-planar coordination originates from the elongation octahedron along the out-of-plane direction [32,34]. The elongation of the octahedron belongs to the Jahn-Teller (JT) distortion, which is attributed to the degeneracy of the *d*-electron state. The JT distortion degree is estimated by a distortive degree parameter $\Delta_d = 1/6 \sum_{i-1}^{6}[(R_i - \bar{R})/\bar{R}]^2$, where $R_i$ and $\bar{R}$ denote individual and average values of the Pd-*X* bonds, respectively [63]. The out-of-plane chalcogen *X* ions gradually approach the coordinated central Pd ions with apparent displacements, which leads to a rapid decrease of the distortive degree ($\Delta_d$) as a function of pressure below 3 GPa (as shown in **Figure S1** [60]). The interlayer Pd-*X* distances are about 3.3 (Pd-S) and 3.2 (Pd-Se) Å at ambient conditions, which are on the verge of the vdW bond (3.1 ~ 5.0 Å) [34]. The interlayer Pd-*X* distances shrink quickly along with compression and go below the threshold value of vdW interactions when the pressure increasing up to 3 GPa. Further compression leads to the collapse of interlayer spacing and the annihilation of vdW interactions. Consequently, the distortive degree ($\Delta_d$) of the Pd$X_6$ octahedron is sharply suppressed at 3.5 GPa, and the orthorhombic lattice changes to a cubic one up to 4 GPa. Meanwhile, the coordination polyhedron transforms from a square-plane into an almost normal octahedron [**Figures 2 (e) and (f)**], as demonstrated by the tiny $\Delta_d$ in **Figure S1** [60]. The electron localization function (ELF) reveals the changes of the interlayer bonding interactions [**Figuer 2 (g)**]. At 0 ~ 3 GPa, the ELF is almost zero in the interlayer spacing, where no covalent bond exists. Above 3.5 GPa, the interlayer charge density significantly increases, and a covalent interaction gradually appears between the metal Pd and the chalcogen *X* ions in the adjacent layer. The dynamic stability of the cubic phase at 4 GPa is examined by the phonon spectrum (**Figures S2**) [60]. The phonon spectrum obtained by dDsC dispersion correction method and PBE exchange-correlation functional has an evident imaginary frequency at $\Gamma$ point, yet the imaginary frequency is completely eliminated and there are no soft acoustic branches in the whole path by switching to the local-density approximation (LDA) functional, indicating that the imaginary frequency is not caused by structural instability [16]. Therefore, the compression almost eliminates the JT distortion in PdSSe and renders the cubic structure stable under high pressure. Recent experiments have shown that PdSe$_2$ underwent a structural transformation from layered orthorhombic phase to cubic pyrite-type structure at 6 ~ 10 GPa [33,35], while a similar structural transition occurs at a pressure of 16 GPa for its isostructural PdS$_2$ [32]. Our previous theoretical studies have confirmed the emergence of a stable cubic pyrite structure when the JT distortive degree of PdS$_2$ was absolutely eliminated by pressure [34], whereas the distortive degree of cubic phase PdSSe still retains a certain value due to the different ionic radii of S and Se. Anyway, we can still expect that structural transformations from the layered orthorhombic phase to cubic phase will be realized in PdSSe by high-pressure experiments according to the pressure homology principle [32-35].

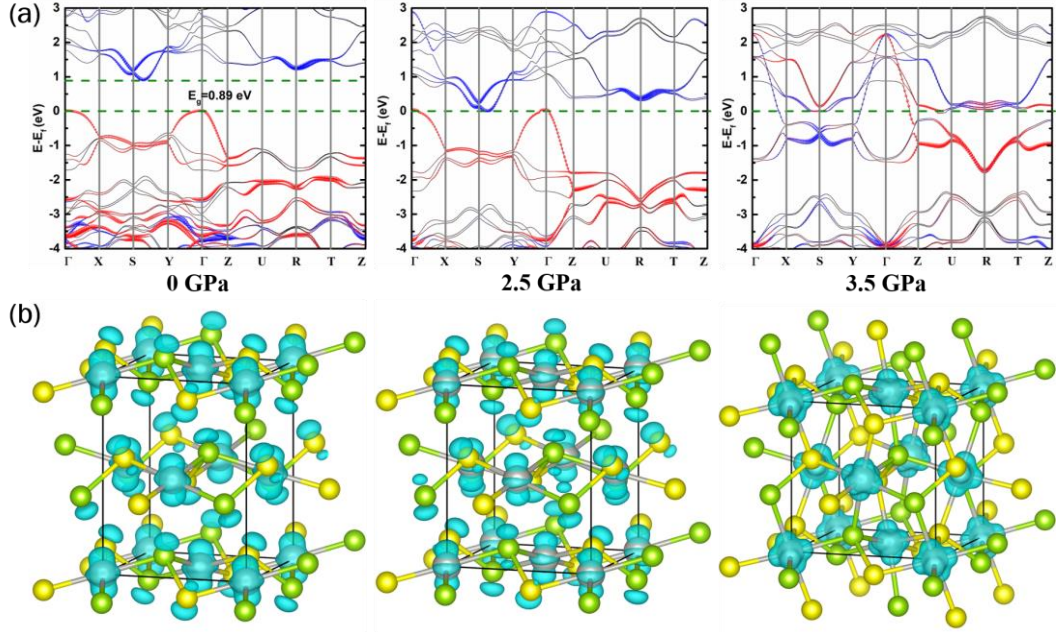

**Figure 3** The electronic structure of PdSSe under different hydrostatic pressures: (a) band structures calculated by HSE06. The contributions of the $d_{x^2-y^2}$ (blue balls) and $d_{z^2}$ (red balls) orbitals are proportional to the size of the colored balls. (b) the corresponding partial charge density. The selected energy range of charge density is set to -1 to 0 eV.

### B. Electronic structure transitions from semiconductor to semimetal under compression

The pressure-induced structural transition is also accompanied by electronic structure variations. Like many semiconducting TMDCs materials [64], as the inlayer distances decrease during compression, the band gaps of PdSSe gradually decrease until complete closure (**Figure S3** [60]). Under ambient pressure, the band structures of all the three polymorphs PdSSe show indirect semiconducting characteristics [50]. The calculated band gaps ~ 0.89 eV are slightly larger than the experimental value of 0.65 eV [10], which may be due to the overestimation of the HSE06 functional (**Figure 3**). In addition, due to the elongated lattice constants in the *c* direction, the calculated band gaps in the present paper are slightly larger than those of our previous work (0.62 ~ 0.78 eV) [50]. This phenomenon indicates that the band gap of PdSSe is extremely sensitive to the interlayer distance, which is like the case of PdSe$_2$ [16]. The band gap of PdSSe closes up at 2.5 GPa. As shown in **Figure 3(a)**, the band dispersion at 2.5 GPa is very similar to that at 0 GPa, but the valence band maximum (VBM) is higher than the conduction band minimum (CBM), displaying a typical characteristic of semimetal. Further increasing pressure up to 3.5 GPa leads to the lattice transform to an intermediate phase with collapsed interlayer spacing. Though the orthorhombic symmetry remains unchanged, the interlayer interactions are dominant by the covalent bond rather than vdW forces. The collapsed interlayer spacing gives rise to a nearly metallic band structure for the intermediate phase. In fact, there are obvious charge gap between the valence bands and conduction bands along most parts of the high-symmetric paths [right column of **Figure 3(a)**], while they only nearly contact with each other along the high-symmetric paths of $\Gamma$-$X$, $\Gamma$-$Y$, and $\Gamma$-$Z$ (see a zoom-in picture of the band structure at 3.5 GPa in **Figure S4** [60]). Therefore, the pressure-induced intermediate phase of PdSSe can be still viewed as semimetal.

Previous resistivity experiments observed semiconductor to metal transitions in PdS$_2$ (PdSe$_2$) at 7 (3) GPa [32,35]. However, the pressure-induced metallization in PdS$_2$ and PdSe$_2$ occurs before the structure transition to the cubic phase. Therefore, the metallization of the intermediate phase in PdS$_2$ and PdSe$_2$ is not related to the cubic structure, but originated from the broadening of the energy bands and filling-changes of the Pd-$d$ orbitals around the Fermi level [32,34]. We propose that the pressure-induced metallization in PdSSe can also be explained by an orbital selective mechanism. As shown in **Figure 3**, the distributions of the Pd 4$d$ orbital near Fermi energy are obviously changed under pressure, which are consistent with the variation of the coordination environments of the Pd atoms. Before the structure changing to the intermediate phase, the $d_{x^2-y^2}$ and $d_{z^2}$ orbitals are separated from each other, and dominant to the CBM and VBM, respectively [34]. The electronic structure transition from semiconductor to semimetal originates from the bandwidth broaden [middle column of **Figure 3(a)**], which can be completely attributed to the increase of the hopping due to the reduced interlayer distance. When the lattice collapses into the intermediate phase, the coordination polyhedron of Pd ion changes from square plane to distorted octahedron. Accordingly, the corresponding energy of $d_{z^2}$ level increases, which results in a reduction of the crystal-field splitting between $d_{x^2-y^2}$ and $d_{z^2}$ orbitals. Mixed-filled $d_{x^2-y^2}$ and $d_{z^2}$ orbitals obviously contribute to the bands around Fermi level [right column of **Figure 3(a)**], implying the orbital selective mechanism is responsible for the metallization [32,34].

The metallization process can be further inspected by the variation of the band structure combined with local charge density analysis. As shown in **Figure 3 (b)**, we can see obvious variations of the occupied orbital along with the increasing pressure. Similar to PdS$_2$ and PdSe$_2$, the Pd ion has a nominal +2 valence state with 4$d^8$ electron configuration in PdSSe. At ambient-pressure conditions, the 4$d$ electrons of Pd$^{2+}$ preferentially occupy lower energy $d_{xz}$, $d_{yz}$, $d_{xy}$, and $d_{z^2}$ orbitals according to the square-planar crystal-field splitting [65]. Therefore, the empty $d_{x^2-y^2}$ orbitals locate above the Fermi level and are separated from the occupied $d_{z^2}$ orbitals by a charge gap. In addition, the dispersions of the VBM along $\Gamma$-$X$ and $\Gamma$-$Y$ directions are highly similar at ambient pressure, which are weaker than the highly dispersive bands along $\Gamma$-$Z$ direction (left column of **Figure 3**), suggesting a strong interlayer coupling interaction [11,12]. Upon compression, the chalcogen ions $X$ from the adjacent layers get close to the central Pd atom, resulting in a distorted octahedral coordination. The uppermost valence bands are broadened along with increasing pressure, and the gradually increasing energy of $d_{z^2}$ orbital leads to a shrinkage of the energy gap between the $d_{x^2-y^2}$ and $d_{z^2}$ orbitals, resulting in semiconductor to semimetal transition (middle column of **Figure 3**). Further increasing the pressure, the distortive degree of octahedra is almost annihilated after the collapse of the interlayer spacing (**Figure S1** [60]), then the interlayer covalent bonds are established. The reductions of crystal-field splitting between $d_{x^2-y^2}$ and $d_{z^2}$ orbitals cause a mixture of the empty $d_{x^2-y^2}$ orbitals with the filled $d_{z^2}$ orbitals, which forms doubly degenerate $e_g$ states ($d_{x^2-y^2}$ and $d_{z^2}$). Under 3.5 GPa, the triply degenerate full-filled $t_{2g}$ states ($d_{xy}$, $d_{xz}$ and $d_{yz}$) form localized narrow bands at lower energy region, which are separated from the half-filled $e_g$ states ($d_{x^2-y^2}$ and $d_{z^2}$) by a big gap of ~ 0.57 eV comparable to the counterpart in PdS$_2$ (0.7 eV) [34]. The half-filled $e_g$ states show similar dispersion characteristics along $\Gamma$-X, $\Gamma$-Y and $\Gamma$-Z directions (**Figure S4** [60]). Especially, the bands crossing Fermi level along $\Gamma$-X and $\Gamma$-Y directions are mainly contributed from the $d_{x^2-y^2}$ orbitals, in contrast to the $d_{z^2}$ orbitals along $\Gamma$-Z direction, which results in a semimetallic band structure in the intermediate phase (right column of **Figure 3(a)** and **Figure S4** [60]). Finally, the intermediate phase transforms into the cubic phase upon pressure increasing to 4 GPa, whose transport properties under high pressure will be discussed in the next subsection.

### C. Topological superconductivity of the high-pressure cubic phase

More interestingly, previous experiments demonstrate that the high-pressure cubic phase PdS$_2$ and PdSe$_2$ show superconductivity and display dome-shaped superconducting diagrams [32,35]. Electron structure analysis of the cubic phase PdS$_2$ and PdSe$_2$ indicates that the superconductivity under high pressure is related to the Dirac states and nodal-line states near the Fermi level [34,35]. As detailed shown in **Figure 4**, the band structure of the high-pressure cubic phase PdSSe highly resembles those of the cubic superconducting PdS$_2$ and PdSe$_2$. The bands crossing Fermi level exhibit excellent metallicity with strong hybridization between the S 3$p$, Se 4$p$ states and Pd 4$d$ states as demonstrated by the projected band structure in **Figure S5(a)**. The parabolic bands around the Fermi level at $\Gamma$ point are considered to be associated with strong electron-phonon coupling and conventional superconductivity. Therefore, we speculate that the transport properties of PdSSe are qualitatively similar to the superconductivity of PdS$_2$ and PdSe$_2$ under high pressure.

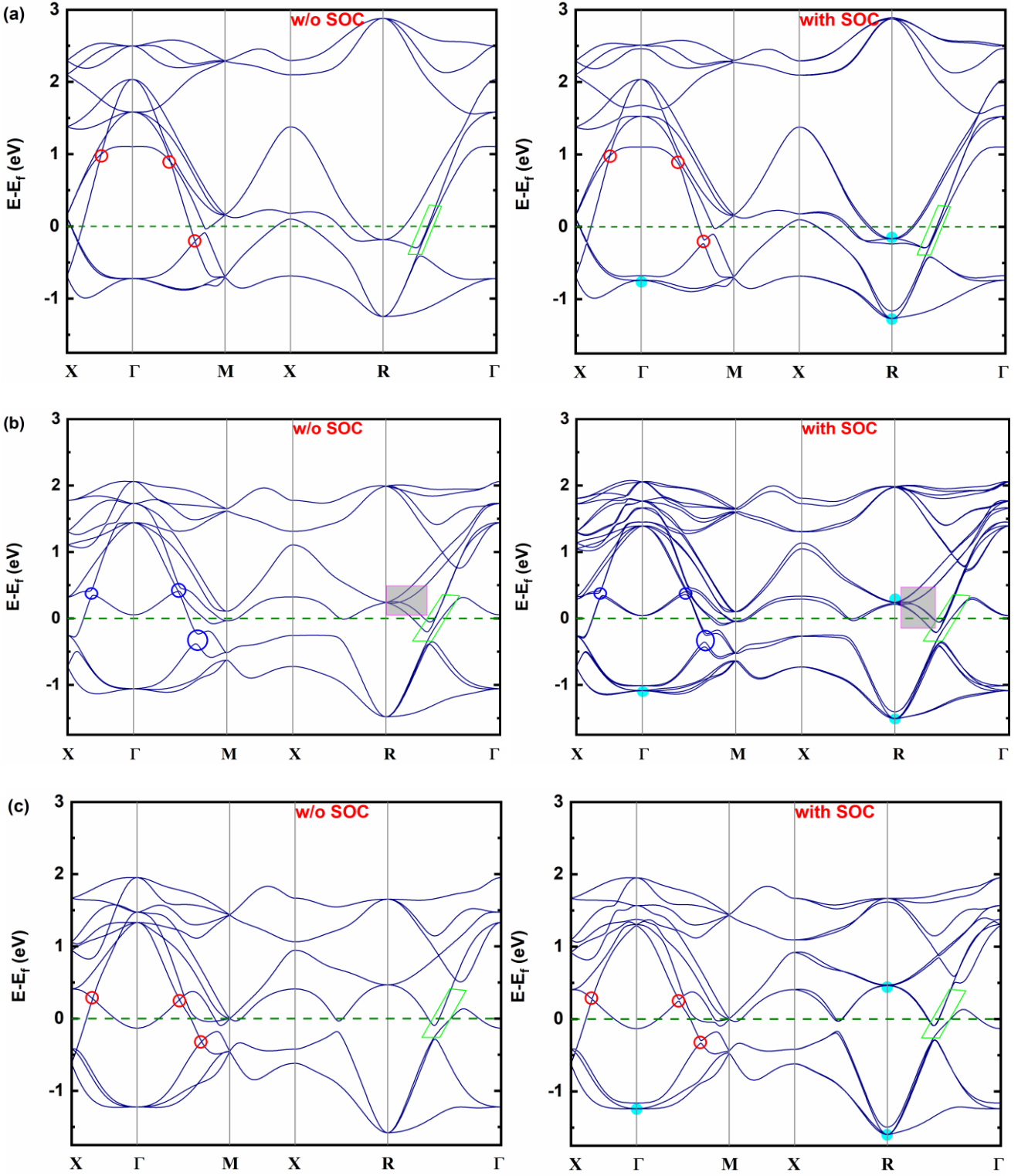

**Figure 4** A comparisons of the band structure of the cubic phase (a) PdS$_2$, (b) PdSSe (*AA* polymorph), and (c) PdSe$_2$ at pressure of 4 GPa without (w/o) or with SOC interactions. The green rhombus illustrates the charge gap along the *Γ*-*R* directions. The Dirac and nodal-line states are marked by the red circles in (a) and (c), while they are broken as indicated by the blue circles in (b). The cyan points mark the fourfold and sixfold degenerate fermions at the *Γ* point and *R* point. The Weyl points along the *Γ*-*R* line are highlighted by the shadow rectangles in (b), which are enlarged

shown in **Figure S6** [60].

However, in contrast to the eightfold degeneracies (considering the spin degeneracy) at the *M* point in PdS$_2$ and PdSe$_2$, bands in PdSSe at *M* point split into pairs of fourfold degenerate bands. In addition, the inversion symmetry breaking furher lifts the degeneracies of the band structure along the high-symmetric directions when the spin-orbit coupling (SOC) interactions are taken into account. Especially, the Dirac states and nodal-line states protected by the glide mirror symmetry of the space group $Pa\bar{3}$ in PdS$_2$ and PdSe$_2$ are broken in PdSSe. These significant changes of the band structure originate from the absence of inversion and mirror-plane symmetries in PdSSe with respect to that of the cubic pyrite-type PdS$_2$ and PdSe$_2$. Therefore, strictly speaking, the structure symmetry (No. 198 space group $P2_13$) of the high-pressure phase PdSSe (*AA* polymorph) belongs to the cubic ullmannite-type structure rather than the cubic pyrite-type structure with the No. 205 space group $Pa\bar{3}$ [66]. As shown in **Figure 4**, the two-fold screw rotation ($\tilde{c}_{2\alpha}$) and three-fold rotation ($C_3$) symmetries of the nonsymmorphic cubic space groups (No. 198 and No. 205) give rise to exotic fourfold and sixfold degenerate fermions at the *Γ* point and *R* point, respectively [67,68]. Particularly, the lowest conduction band and the highest valence bands along high symmetry lines are completely gapped out after considering the SOC interactions for PdSSe, PdS$_2$ and PdSe$_2$, demonstrating typical characteristics of topological semimetal for these compounds, and providing a possible venue for the emergence of nontrivial surface states. We carry out a trial calculation of the symmetry-based indicator and receive a $Z_4$ = 1 for both the cubic pyrite-type PdS$_2$ and PdSe$_2$ (No. 205 space group $Pa\bar{3}$) [69,70], which indicates strongly topological properties awaiting for further detailed investigation. Furthermore, Weyl fermions can arise in the cubic phase PdSSe due to the breaking of inversion symmetry, as demonstrated by the Weyl points along the *Γ*-*R* line [**Figure 4(b)**] induced by the crossing of bands with different $C_3$ eigenvalues (a close view of the band crossing points near the high-symmetric *R* point around the Fermi level is presented in **Figure S6** [60]) [67]. The superconductivity seems to intertwined with these topologically nontrivial states [71].

As shown in **Figure S5(b)**, the Fermi surface of PdSSe consists of a large cubelike pocket at the center of the Brillouin zone, and a network of small electron and hole pockets connected along the edges of the Brillouin zone, which are consistent with the multifold degeneracies. The cubelike Fermi surface with small curvature indicates that the cubic phase PdS$_2$ and PdSSe are likely to produce quantum oscillations with large effective mass [34,72,73]. Furthermore, the nontrivial topological states close to the Fermi level contribute considerable density-of-states (DOS) values at the Fermi level, which are combined with the large Fermi surface, may provide sufficient channels for electron-phonon scattering [74] and favor an enhancement of the electron-phonon interaction [75]. Therefore, since the cubic phase PdSSe features similar dispersion characteristics of the band structure, topological states, and Fermi surface with the superconducting phase $PdS_2$ and $PdSe_2$, we expect similar superconducting behaviors will be observed in PdSSe by high-pressure experiments.

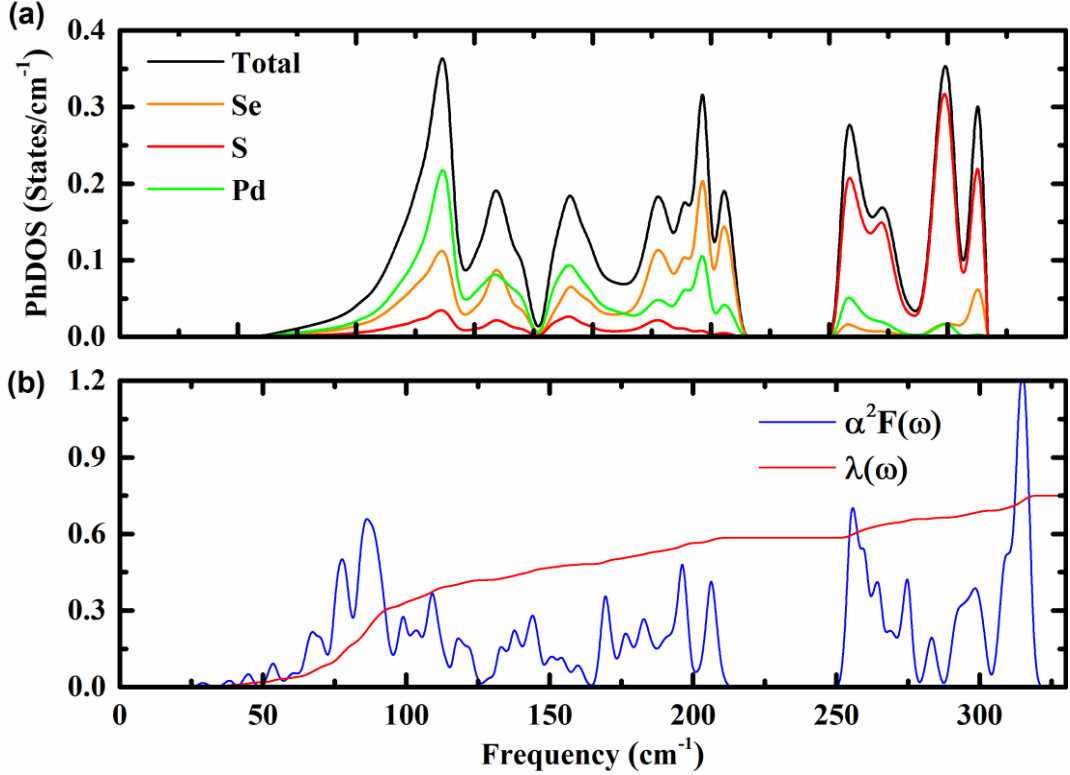

**Figure 5** Calculated (a) phonon density of states (PhDOS), and (b) Eliashberg electron-phonon coupling spectral function $\alpha^2F(\omega)$ and the electron-phonon coupling integral $\lambda(\omega)$ of cubic PdSSe at 4GPa.

In order to confirm the possible superconductivity in cubic phase PdSSe under high pressure, we calculate the electron-phonon coupling constant $\lambda$ and the superconducting transition temperature $T_c$ for PdSSe by the Allen-Dynes-modified McMillan formula [76]:

$$T_c = \frac{\omega_{log}}{1.2}\exp[-\frac{1.04(1+\lambda)}{\lambda-\mu^*-0.62\lambda\mu^*}], \quad (1)$$

where the Coulomb pseudopotential $\mu^*$ is set to 0.1, and the electron-phonon coupling constant $\lambda$ is defined as

$$\lambda = 2\int_0^\infty \frac{\alpha^2F(\omega)}{\omega}d\omega, \quad (2)$$

and $\omega_{log}$ is the logarithmic average frequency:

$$\omega_{log} = \exp[\frac{2}{\lambda}\int \frac{log\omega}{\omega}\alpha^2F(\omega)\,d\omega], \quad (3)$$

where $\alpha^2F(\omega)$ is the Eliashberg spectral function defined as

$$\alpha^2F(\omega) = \frac{1}{2\pi N(E_\text{f})}\sum_q \delta(\omega - \omega_q)\frac{\gamma_q}{\hbar\omega_q}, \qquad (4)$$

where $N(E_\text{f})$ is the DOS at the Fermi level and $\gamma_q$ is the phonon linewidth. The phonon spectrum of cubic PdSSe at 4 GPa does not show any imaginary frequencies and soft acoustic branches (**Figure S2** [60]), demonstrating the dynamical stability of the cubic phase PdSSe under high pressure. Correspondingly, the phonon DOS (PhDOS) displays three regions separated by two gaps (**Figure 5**). The heaviest Pd atoms predominantly contribute to the low-frequency acoustic branches, while the lighter Se atoms contribute to the middle-frequency optical branches. The contributions of the lightest S atoms are very tiny in the low and middle-frequency region, and predominantly dominate the high-frequency optical branches. The broad distribution of electron-phonon interactions is reflected in the calculated Eliashberg spectral function $\alpha^2F(\omega)$ [77]. Calculated results show that there are strong electron-phonon interactions in the pressure-induced cubic phase PdSSe. The electron-phonon coupling constant $\lambda$ of the cubic phase PdSSe at 4 GPa is 0.75, and the corresponding superconducting transition temperature $T_c$ is 7.5 K. Therefore, the theoretically calculated $T_c$ of cubic phase PdSSe at 4 GPa is comparable to the $T_c$ of 4.9 (7.6) K for the cubic pyrite phase $PdS_2$ ($PdSe_2$) at 4 (10) GPa [34]. Furthermore, the $T_c$ of the cubic pyrite phase $PdS_2$ and $PdSe_2$ shows dome-shaped dependence on pressure, which increases up to maximum values of 8.0 and 13.1 K at 37.4 and 23 GPa, respectively [32,35]. We examine the stability and superconductivity of the cubic phase PdSSe under ambient conditions. As shown in **Figure 6**, the phonon band structure without any imaginary frequencies indicates that the cubic PdSSe is dynamically stable at 0 GPa. The corresponding PhDOS almost display identical characteristics to that at 4 GPa. The calculated electron-phonon coupling constant $\lambda$ is 0.68 and the corresponding superconducting transition temperature $T_c$ is 5.8 K, which is lower than the one at 4 GPa. The increasing trend of the theoretically calculated superconducting temperature from 6.1 K (at 0 GPa) to 7.5 K (at 4 GPa) is in line with the experimental observations in cubic pyrite phase $PdS_2$ and $PdSe_2$ [32,35]. Therefore, the pressure-induced evolutions of the crystal structure and electronic transport properties of PdSSe are qualitatively similar to that of $PdS_2$ and $PdSe_2$.

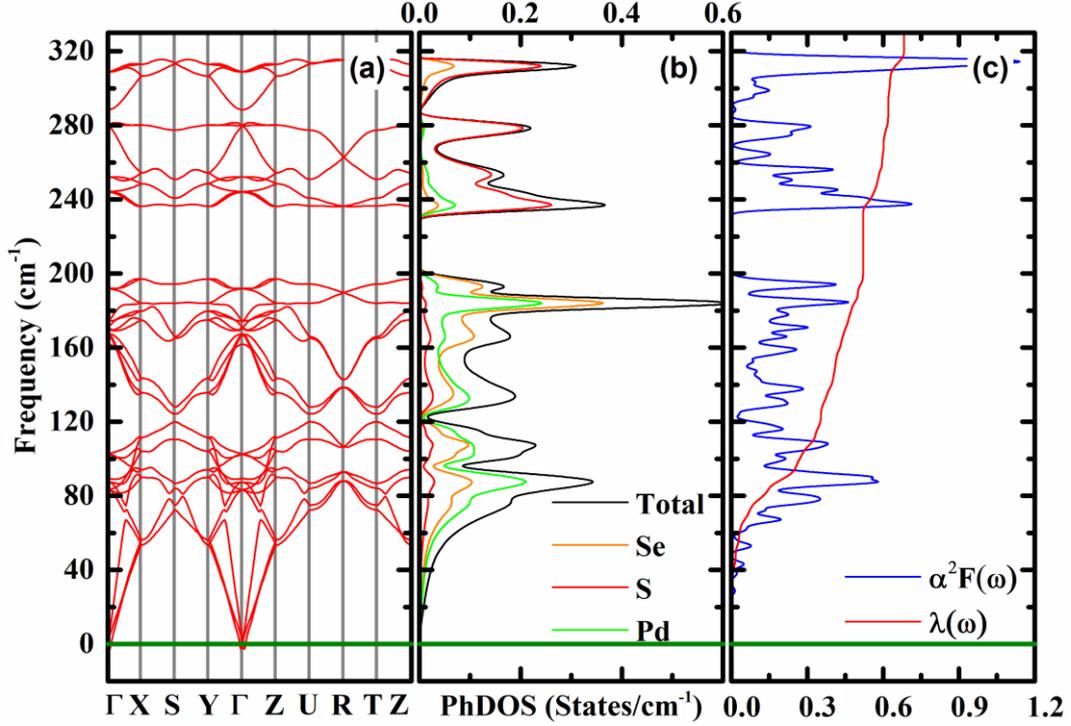

**Figure 6** Calculated (a) phonon spectra of the cubic PdSSe at 0 GPa, (b) the corresponding PhDOS, and (c) the Eliashberg electron-phonon coupling spectral function $\alpha^2F(\omega)$ and the electron-phonon coupling integral $\lambda(\omega)$.

Previous experiments indicate that the superconductivity shows a striking correlation with the structural instability of the chalcogen dumbbells in cubic pyrite phase PdSe$_2$ [35]. However, despite showing similar Se-Se bond softening under pressure, the isostructural and isoelectronic pyrite-phase NiSe$_2$ remains nonsuperconducting up to 50 GPa [32]. Moreover, high-pressure Raman spectroscopy and theoretical calculations do not observe any anomalous softening of the S-S bonding in the isomorphous cubic pyrite phase PdS$_2$ [32,34]. The symmetric chalcogen Se-Se (S-S) dumbbells in the cubic PdSe$_2$ (PdS$_2$) have been broken and replaced by the S-Se bond in the cubic phase PdSSe. And our theoretical calculations reveal that the S-Se bond lengths in cubic phase PdSSe are nearly independent on pressure [**Figure 2 (b)**], similar to the rigid S-S bond observed in the cubic pyrite phase PdS$_2$. Therefore, the structural instability of the chalcogen dumbbells is not directly correlated with the emergence of superconductivity in the pyrite-type compounds. The topological nontrivial states around the Fermi level and strong electron-phonon coupling might play a crucial role in triggering the superconductivity in the cubic phase PdSSe under high pressure, waiting for further experimental studies to clarify.

## IV. CONCLUSIONS

In conclusion, we have investigated the transformations of crystal structure and the concomitant evolutions of transport properties in PdSSe under hydrostatic pressure. PdSSe shows similar structural transition and transport behavior to that observed in its sister compounds of $PdS_2$ and $PdSe_2$. Pressure-induced structural transitions from layered orthorhombic phase to cubic phase through an intermediate phase have been theoretically predicted in PdSSe, which are accompanied by a series of electronic structure transitions from semiconductor to semimetal to superconductor. Electronic structure and electron-phonon coupling calculations indicated that the superconductivity can be realized in the high-pressure cubic phase, which originated from the strong electron-phonon coupling associated with topological nontrivial states. The present study not only reveals the pressure-induced structural transformations and semiconductor-semimetal-superconductor transitions in PdSSe, but also provides new insights into the metallization and topological superconducting properties of the TMDCs compounds under high pressure, which will attract further experimental and theoretical explorations of interesting physics in TMDCs.

## ACKNOWLEDGMENTS

We would like to thank the fruitful discussion with Dr. Xiaoming Zhang and Xiaobo Ma. X. M and J. L are sponsored by the National Natural Science Foundation of China (No. 11864008, 12134018 and 11921004) and Guangxi Natural Science Foundation (No. 2018AD19200 and 2019GXNSFGA245006). C. A. is supported by the Foundation for Polish Science through the International Research Agendas program co-financed by the European Union within the Smart Growth Operational Programme.